\documentclass{ws-procs9x6}

\begin{document}

\title{Cosmological Parameters: Fashion and Facts}

\author{A. BLANCHARD}

\address{Laboratoire d'astrophysique de l'OMP, CNRS, UPS\\
14, Av. E.Belin, 31~400 Toulouse, FRANCE\\ 
E-mail: Alain.Blanchard@ast.obs-mip.fr}


\maketitle

\abstracts{
We are at a specific period of modern cosmology,
during which the large increase of the amount of data relevant to cosmology, 
as well as their increasing accuracy,
leads to the idea that the determination of cosmological parameters
has been achieved with a rather good precision, may be of the order of 10\%.
There is a large consensus around the so-called concordance model.
Indeed  this model does  fit an impressive set of
independent data, the most impressive been: CMB $C_l$ curve, most current 
matter density estimations, 
Hubble constant estimation from HST, apparent acceleration of the Universe,
good matching of the power spectrum of matter fluctuations. 
However, the necessary introduction of a non zero
cosmological constant is an extraordinary new mystery for physics,
or more exactly the come back of one of the ghost of modern physics
since its introduction by Einstein. Here,
I would like to emphasize that some results are established beyond
reasonable doubt, like the (nearly) flatness of the universe and the existence 
of a dark non-baryonic component of the Universe.
But also that the evidence for a cosmological constant may not be as strong
as needed to be considered as established beyond doubt.
In this respect, I will argue that an Einstein-De Sitter universe
might still be a viable option.  Some observations do not 
fit the concordance picture, but they are generally considered as not to be
taken into account. I  discuss several of 
the claimed observational evidences supporting the concordance model,
and will focus more specifically on 
the observational properties  of clusters which offer powerful constraints
on various quantities of cosmological interest. They are particularly
 interesting
 in constraining the cosmological density parameter, nicely complementing
the CMB result and the supernova probe. While early estimations were
based on the of the $M/L$ ratio, i.e. a local indirect measure
of the mean density which needs an extrapolation over several orders of
magnitude, new  tests have been proposed during the last ten years which
are global
in nature. Here, I will briefly discuss three of them: 1) the evolution of the
abundance of clusters with redshift 2) the baryon fraction measured in local
clusters 3) apparent evolution of the baryon fraction with redshift.
I will show that these three independent tests lead to high matter density
for the Universe in the range $0.6-1.$. I therefore conclude that the 
dominance  of  vacuum  to the various density contributions to the Universe
is presently an interesting and fascinating possibility, but it is still
premature to consider it as an established scientific fact.  }

\section{Introduction: the contents of the Universe}

Cosmology is a rather young field still undergoing a very fast evolution.
Twenty years ago the nature of the microwave background was still
a matter of debate, although it was generally believed that
the origin was mainly from the Big Bang, the exact shape of
the spectrum was still uncertain. The measurement of the spectrum
 by COBE, which was concomitant to the Gush et al. measurement (1990),
  showing that the spectrum was a nearly perfect blackbody
has been a fundamental result in modern cosmology, by establishing in a 
definitive way one of the most critical prediction of the 
standard hot Big Bang picture. 
The determination of cosmological parameters is a central question in modern
cosmology and it has become more central after the 
next fundamental result established by COBE: the first robust 
detection of the CMB fluctuations, nearly thirty years after the prediction 
of their presence (Sachs and Wolf, 1967).
This detection  has opened a new area
with the perspective of reaching high ``precision  cosmology''. However,
it is also important to mention the fact that the Inflation 
paradigm (Guth, 1981)
has represented an enormous attraction for theorist towards the field 
of cosmology, opening the perspective of properly testing high energy 
physics from cosmological data, while such a physics will probably remain
 largely 
unaccessible from laboratory experiment. Even if the data from the CMB 
fluctuations were not taken at face values as a proof of inflation
the need for new physics appear very strongly (it is interesting
to mention  that the 
origin of the asymmetry  between matter and antimatter was 
a fundamental 
problem which solution involves  physics of the very early universe).

Moreover, the presence for 
non-baryonic dark matter can be now considered as a well established fact of
modern physics. This was far from being obvious twenty years ago. By
 present days the abundances of light elements is well 
constrained by observations,
consistent with a restricted  range of baryon  abundance (O'Meara et al., 2001):
\begin{equation}
\Omega_{baryons} = 0.02 h^{-2} \pm 0.002
\label{eq:obbn}
\end{equation}
where $h$ is the Hubble constant in unit of $100$ km/s/Mpc.
 The above baryon abundance is in full agreement with what can be inferred 
from
CMB (Le Dour et al., 2000;  Beno\^{\i}t et al., 2002b). 
There are differences in matter density 
estimations, but nearly all
 of them lead to a cosmological 
density parameter in the range [0.2--1.], and therefore  those estimates 
imply the presence of a non-baryonic component of the density of the universe.
An other implication is that most baryons are dark: the amount of baryons
seen in the Universe is mainly in form of stars:
\begin{equation}
\Omega_{\rm stars} = 0.003-0.010
\label{eq:stars}
\end{equation}
much less than predicted by primordial nucleosynthesis. 
This picture, the presence of two dark components in the Universe, has gained
considerable strength in the last twenty years, first of all because the above
 numbers have gained in robustness. However, it is now believed 
that a third dark constituent has been discovered:  the dark energy.  

\section{Observions and cosmological parameters}

\subsection{What the CMB does actually tell us?}

\subsubsection{The curvature of space}

The
detection of fluctuations on small angular scale, mainly by the Saskatoon 
experiment (Netterfield et al, 1995)  more than 7 years ago
provided a first convincing piece of evidence for a nearly  flat universe
(Linewaever et al., 1997; Hancock et al., 1998; Lineweaver and Barbosa, 1998),
 or more precisely
evidence against open models which were currently favored at that time. This conclusion
 is now firmly
established thanks to high precision  recent measurements including  those of 
Boomerang,
Maxima and DASI (de Bernardis et al;, 2000; Hanany et al., 2000; Halverson et al., 2002): open models 
are now 
entirely ruled out: $\Omega_t > 0.92$ at 99\% C.L., it should be noticed  
that upper limit on $\Omega_t$ are less stringent, $\Omega_t < 1.5$ at 99\% 
C.L., unless one add some prior,
for instance  
on the Hubble constant. 

\begin{figure}[ht]
\centerline{\epsfxsize=3.in\epsfbox{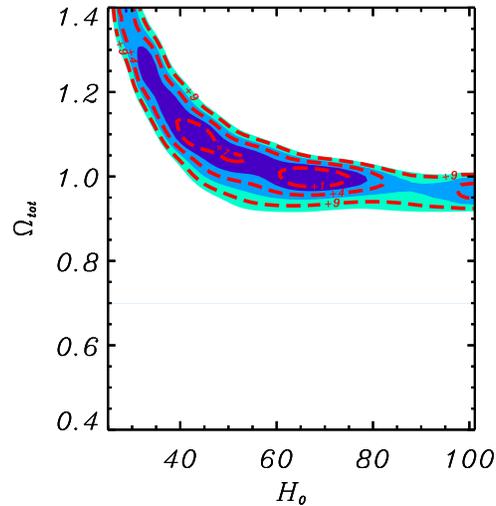}}   
\caption{On this picture the likelihood contours from the CMB
constraints are given: dashed lines, when projected provide the 
68\%, 95\%, 99\% confidence intervals, while shade area correspond to 
the contours on two parameters. The likelihood is maximized on the 
other parameters. This diagram illustrates several aspects of
constraints that can be obtained from CMB: flatness of the universe 
follows from the fact that $\Omega_t > 0.92$ at 99\% C.L., but $H_0$ is
very poorly  constrained. Indeed CMB allows to severely tighten the 
model parameters space, but can leave us with  indetermination on 
specific individual parameter because of degeneracies.
See Douspis et al., (2000) for further details. 
\label{fig:flatness}}
\end{figure}

The most recent measurements of CMB anisotropies, including those 
obtained after
this conference by Archeops (Beno\^{\i}t et al., 2002a), 
provide a remarkable success of the theory: the detailed shape of the 
angular power spectrum of the fluctuations, the theoretical predictions
of the $C_l$ curve, 
is in excellent agreement with the 
 observational data. This success gives 
confidence in the robustness of conclusions drawn from such analyzes, while
alternative theories, like cosmological defects (Durrer et al, 2002) are almost entirely ruled 
out as a possible primary source of the fluctuations in the C.M.B.
This gives strong support for theories of  structures formation based
the gravitational growth of initial passive fluctuations, the gravitational 
instability scenario, a picture sketched  nearly seventy years ago 
by G. Lema\^{\i}tre (1933). At the same time this implies 
that conclusions on cosmological parameters from CMB have to be considered as robust:
the spectacular  conclusion that the universe is nearly flat
space\footnote{It is sometimes believed that a space cannot be ``nearly'' flat,
 because mathematically space is flat or not. This is not true in Cosmology
where there is a natural scale which is $c/H_0$. Stating that the Universe is
nearly flat means that its curvature radius $R_c$ is much larger than this
scale.} is a major scientific result of modern science which is certainly robust and is very likely to  remain
as one of the greatest advance of modern Cosmology.

\subsubsection{A strong test of General Relativity}

Contrarily to a common conception, General Relativity (GR) is weakly tested on cosmological scales:
the expansion of the Universe can be described in a Newtonian approach, while 
departure from the linear Hubble diagram are weak, and therefore does not 
provide strong test of GR. Actually the observed 
Hubble diagram is used to fit the amplitude of the cosmological constant, i.e.
ones assumes (a non-standard version of!) GR and fits one of the parameter, therefore this does not constitute a test of the theory. However, the $C_l$ curve of  CMB fluctuations
 provides an interesting test of GR on such scales: 
the angular distance to the CMB
accordingly to RG is such that:
\begin{equation}
D_{\rm ang} \sim \frac{1.}{300} c(t_0-t_{\rm lss})
\end{equation}
($t_0$ being the present age of the universe, and $t_{\rm lss}$ the age of
the universe at the last scattering surface (lss) from where the $C-l$ curve 
is produced). This means  that the angular distance to the CMB is of the same order than 
the one to the Virgo cluster! Therefore the $C_l$ curve can be obtained only within a theory where photons
trajectories are essentially those predicted by GR.

\subsection{Is the Universe accelerating ?}

It is often mentioned that the present day data on the CMB excluded 
a model without a cosmological constant. Given the present-day 
quality of the data, and the anticipated accuracy one can hope from 
satellite experiments, this is a crucial issue. Actually, what's happen 
is that
an Einstein-de Sitter is at the boarder  of the $3-\sigma$ contour in 
likelihood analysis.  But this is not sufficient to claim that the model 
is excluded at $3-\sigma$! Actually, a model without a cosmological constant
provides a very acceptable fit to the data in term of a goodness of fit.
Therefore, {\it CMB data do not request a non-zero cosmological constant}.

\begin{figure}[ht]
\centerline{\epsfxsize=2.5in\epsfbox{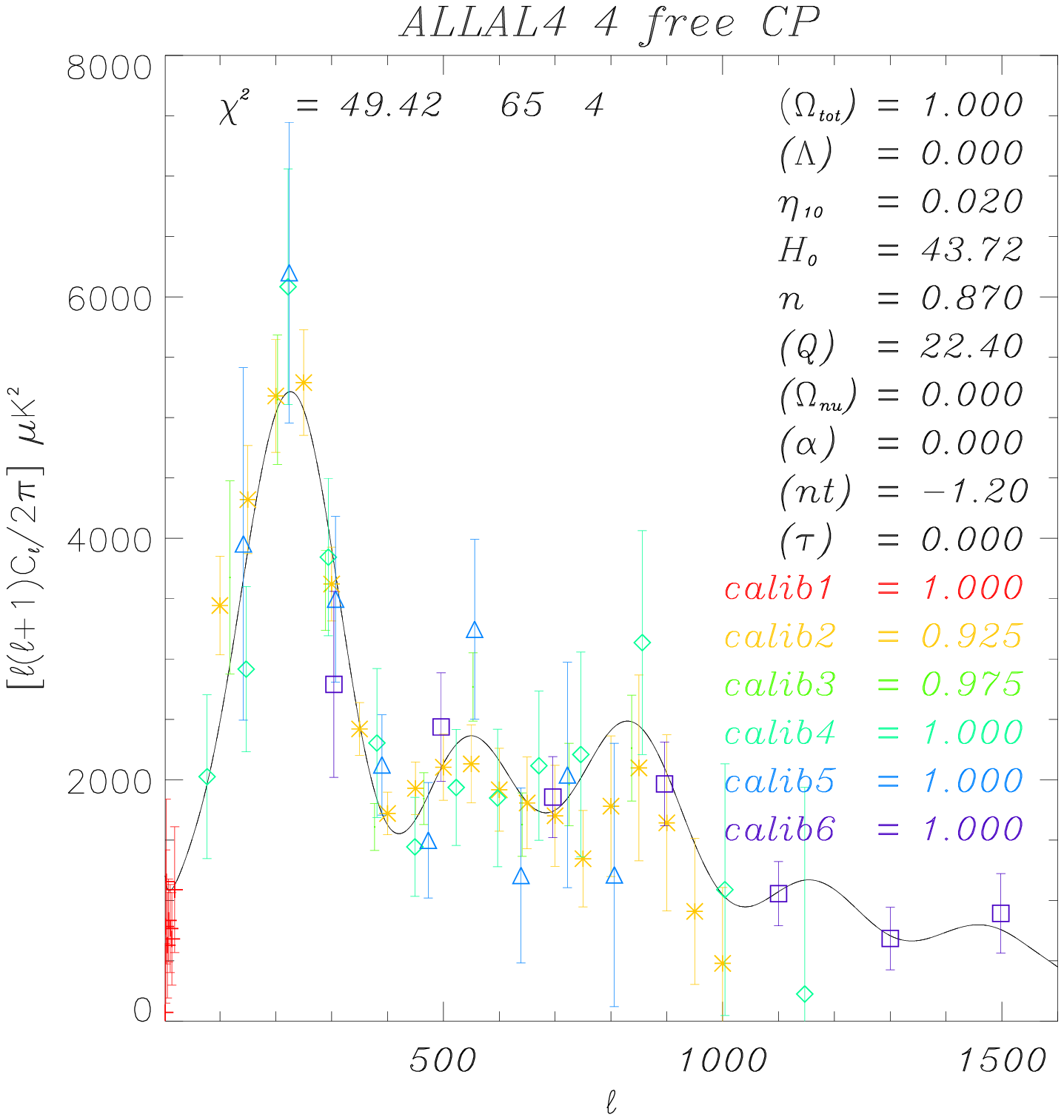}\epsfxsize=2.5in\epsfbox{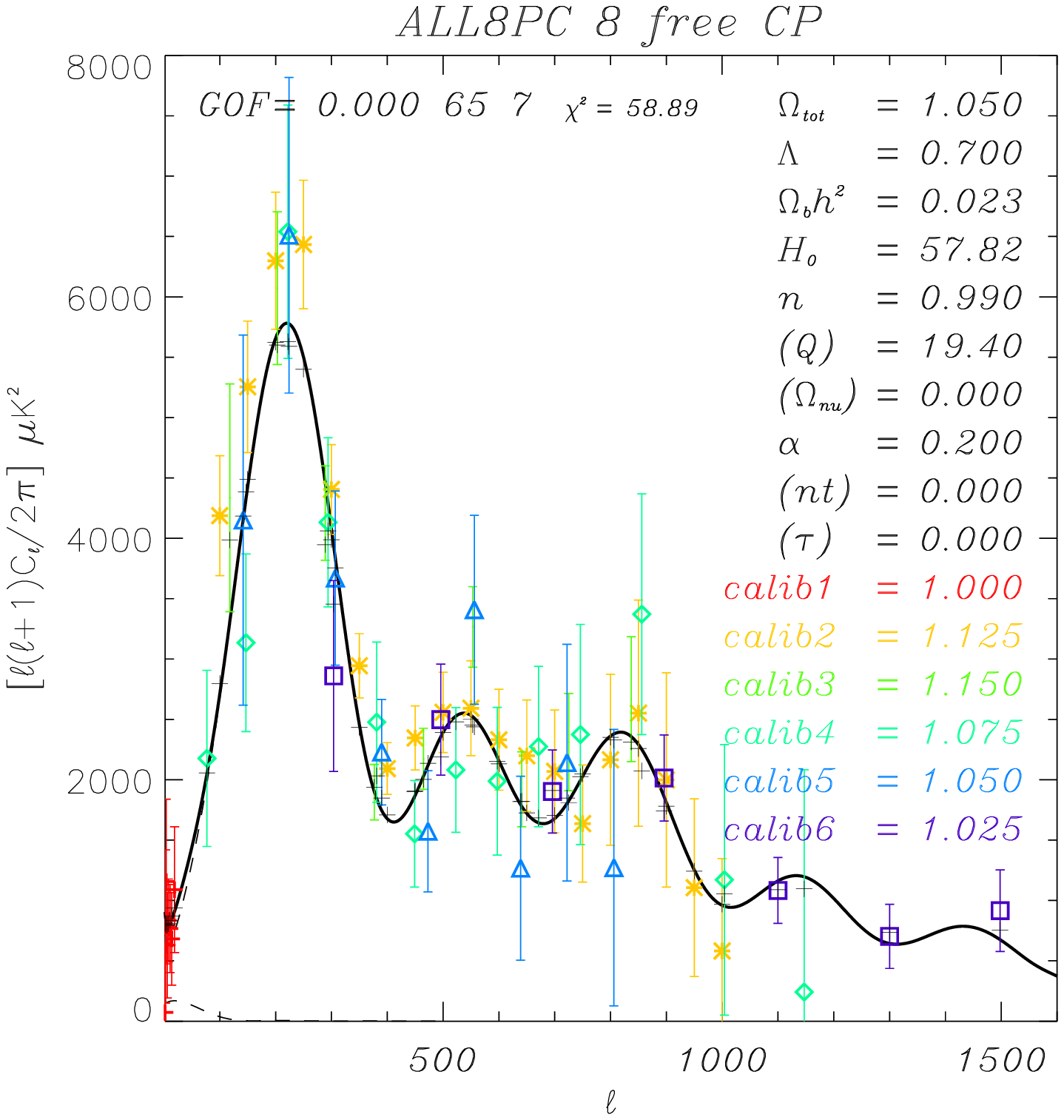}}   
\caption{An example of an acceptable model to the CMB data without inclusion 
of a cosmological constant (left) The Hubble constant has been taken to a low value 
of
44 km/s/Mpc. An exemple of optimal model is shown, corresponding to the concordance model.
Both provide an acceptable fit in term of goodness of fit. Courtesy of M. Douspis. 
\label{fig:flatness2}}
\end{figure}

The possible detection of a cosmological constant from distant supernovae
has brought the essential piece of evidence largely comforting the
so-called concordance model:  the apparent luminosity of distant supernovae
now appears  fainter, i.e. at larger distance, than expected in any decelerating 
universe (Riess et al, 1998; Perlmutter et al., 1999) and can therefore  be 
explained
only within an accelerating universe (under the assumption of 
standard candle). Indeed a CDM model in a flat universe dominated
by a cosmological constant is in  impressive agreement with most of existing
 data: such a model is consistent with the HST measurement of the Hubble
constant, the age of the Universe, the power spectrum and the 
amplitude of matter fluctuations as
measured by  clusters abundance and weak lensing on large scale,
as well as most current measurements of the mass content on
small scales obtained by various technics. The concordance model offers 
therefore a remarkable success for the CDM theory, but at the expense of 
the introduction of a non-zero cosmological constant.

\subsection{Some reasons for caution}

Despite the above impressive set of agreements cited above, one should keep 
an open mind. The question of the age of the Universe is not an issue: models 
fitting adequately the $C_l$ curve leads to similar ages, consistent with existing constraint. For instance, 
the model drawn
on figure 2 (left side) has an age of 15 Gyr, well consistent with age 
estimates (actually a model with $t_0 \sim 10$ Gyr should probably not be securely 
rejected on this basis). Identically the amplitude of matter fluctuations
 on small scales is sometimes claimed to be inconsistent with a high density 
universe, while there is actually a  
degeneracy between this amplitude and the matter density parameter $\Omega_m$.
Very often, authors implicitly refer to the standard CDM scenario 
($\Omega_m = 1$, $h \sim 0.5$, $n = 1$). Actually this simplest CDM model is 
known to be ruled out from several different arguments, but there exists 
also different way one can imagine the spectrum to be modified in order
to match the data (an example is a possible contribution of hot dark matter
of the order of 20\%).

A high density universe is actually inconsistent (because of the age problem)
with 
value of the Hubble constant as high as those found by the HST. 
However, the HST measurement of the Hubble constant has been questioned 
(Arp, 2002).
In order to illustrate the argument I show the figure given by Arp, which is 
claimed to represent the Hubble diagram from the HST data. Clearly, a 
firm conclusion on the Hubble constant 
from this data seems difficult and actually Arp claims
that data can favor $H_0 \sim 55$ km/s/Mpc. 
An other doubt on the Hubble constant  
comes from the Sunyaev-Zeldovich measurements: in a recent review 
Carlstrom et al.  (2002)
found that the best value slightly depends on the cosmology, but that 
in an Einstein-de Sitter 
model one finds an average values of $H_0 \sim 55$ km/s/Mpc, furthermore 
given that 
such determination suffers from possible 
clumping of the gas (Mathiesen et al., 1999; see below), 
the actual value could be 25\% less! \\

Let us now examine observational direct evidences for or against a non-zero
 cosmological constant. Distant SNIa are observed to be fainter than expected
(in a non-accelerating universe) given their redshift, indicating very directly that the universe is 
accelerating should they be standard candles. The signal is of the order of 0.3 magnitude (compared to an
Einstein--de Sitter universe).  It is important to realize that several
astrophysical effects of the same order are already existing, and that 
their actual amplitude might be difficult to properly evaluated. 
Rowan-Robinson 
(2002) argued for instance that the dust correction might have been 
underestimated in high redshift SNIa,
while such a correction is of the same order of the signal. Identically, the 
K-correction that has to be applied to high redshift supernovae is large (in 
the range 0.5--1. mag) and is estimated from zero redshift spectral templates;
one can therefore worry whether some shift in the  zero-point  would not  
remain from the actual spectrum, with an amplitude  larger than the assumed 
uncertainty (2\%). Identically, the progenitor population at redshift 0.5
is likely to be physically different from the progenitors of local SNIa
(age, mass, metalicity). Consequence on luminosity are largely unknown.\\

Finally it is worth noticing that the first 7 distant SNIa which were analyzed
conducted to conclude  to the rejection of a  value 
of $\lambda$ as large as  0.7:  $\lambda < 0.51$ (95\%) 
(Perlmutter et al., 1997). Several arguments have been used in 
the past or recently to set upper limit on  a dominant contribution of
  $\lambda$ (Maoz et al., 1993; Kochanek, 1996; Boughn et al., 2002).
There is therefore a number of arguments for caution:

\begin{figure}[ht]
\centerline{\epsfxsize=2.5in\epsfbox{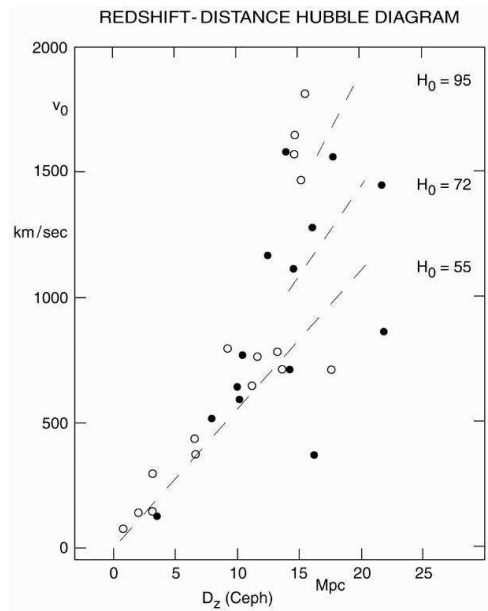}}   
\caption{ Hubble diagram from HST cepheids according to Arp (2002). 
Clearly the derivation of a value from this data set is uncertain.
But a value of 55 km/s/Mpc seems as least as 
adequate as the HST finding (72 km/s/Mpc). 
\label{fig:arp}}
\end{figure}

\noindent 1) SNIa
measurements provide the {\bf single} direct evidence for a cosmological
constant,\\
2) most measurements of $\Omega_m$ are local in nature (mostly inferred
from
clusters),\\
3) some upper limits have been published on $\lambda$ which do not agrre with recent measurements ,\\
 4) a non-zero cosmological constant is an extraordinary new result
in physics and therefore deserves extraordinary piece of evidence.\\
 
Before the existence of the cosmological constant can be considered as
scientifically established, it is probably necessary  to reinforce evidence
for the convergence model by obtaining further direct evidence for a
cosmological constant. Because there exist degeneracies in parameters 
determination
with the CMB, even the Planck experiment will not allow to break 
these degeneracies. It is therefore necessary to 
use  tests which provide complementary information. The data provided by 
the distant SNIa satisfies well this requirement. As it is difficult to think 
of 
a new test measuring directly the presence of a cosmological 
constant, the best approach is  
probably still to try to have reliable estimates of the 
matter density from global technics. In this respect, clusters are 
probably the most powerful tool, as  they provide several  major
roads to measure the density matter of the Universe and
which exploration is still in its infancy. Here I will concentrate
on this perspective.\\
 
\section{The mean density of the Universe from Clusters}
 
The classical way to use clusters to constraint the
average matter density in  to try to obtain a direct measurement of the
local density. This is the principle of $M/L$ test. Because the matter
content of the universe is essentially in a dark
form, we do not have direct measurement of the mass content even 
at the local level.
This is the reason why in practice we rely on a two-steps procedure: first the
average luminosity density of the universe is estimated from galaxy samples,
this quantity is now relatively well known thanks to the large redshift
surveys like the SLOAN or the 2dF (although difference of the order of 50\%
might still exist); the second ingredient is the value  of the $M/L$
 ratio obtained from data on clusters (total luminosity and 
mass estimations). There might be
a factor of two of uncertainty in this quantity. For instance Roussel et al.
(2000) found that the average $M/L$ could be as large as $750h$ when the
mass--temperature relation for clusters is normalized from
 numerical simulations of
Bryan and Norman (1998), while values twice smaller are currently obtained
by other technics. The basic principle for estimating the average
density of the Universe is then to write:  
$$
\rho_m = \rho_l \times M/L
$$
However, it should be realized that the volume  occupied
by clusters  is a tiny fraction  of the total volume of
universe, of the order of $10^{-5}$. The application
of the $M/L$ relies therefore on an  extrapolation over
 $10^5$ in volume!


\begin{figure}[ht]
\centerline{\epsfxsize=4.5in\epsfbox{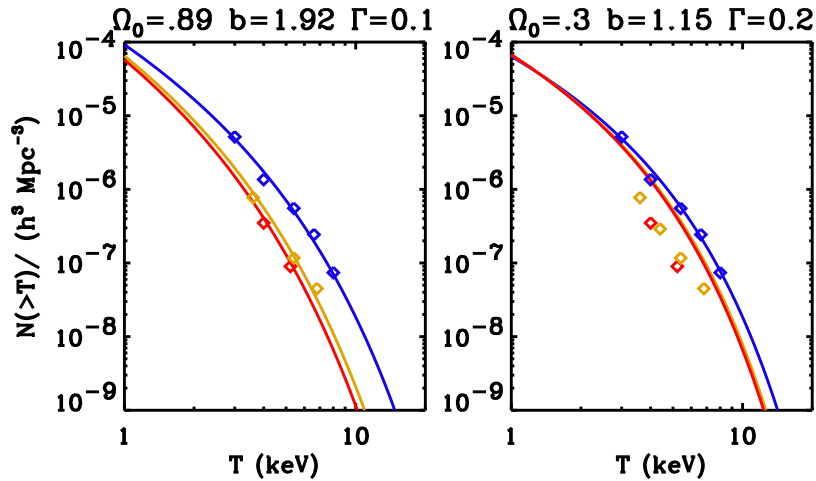}}   
\caption{These plots illustrate the power of the
 cosmological test of the evolution of the abundance of X-ray clusters:
the TDF (temperature distribution function) has been normalized to present day abundance (blue -- dark grey -- lines). The
abundance of local clusters is given by the blue (dark grey) symbols  (Blanchard et al., 2000).
Present abundance allows one to set the normalization and the slope of the
spectrum of primordial fluctuations on clusters scale (which is $\Omega_m$
dependant). The
evolution with redshift is much
faster in a high matter density universe (left panel, $\Omega_m = 0.89$) than
in a low density universe
(right panel, $\Omega_m = 0.3$): $z = 0.33$ (yellow -- light)  the difference
is already of the order of $3$ or larger.   It is relatively insensitive to
the cosmological constant. We also give our estimate of the local TDF  (blue
symbols) derived by Blanchard et al. (2000), as well as our estimate of the
TDF
at $z = 0.33$ (yellow symbols-- light grey).  Also are given for comparison data (Henry
2000) and models predictions at $z = 0.38$ (red -- dark grey -- symbols and
lines). On the left panel, the  best  model is
obtained by fitting  simultaneously  local clusters  and   clusters  at
$z = 0.33$
leading to a best value of $\Omega_m$ of 0.89 (flat universe). The right panel illustrates the
fact that an flat low  density universe  $\Omega_m = 0.3 $ which fits well
local data
does not fit the high redshift data properly at all.
\label{fig:test}}
\end{figure}

 \section{New global tests}
  
In order to have a reliable estimation of the mass density of the universe,
it is vital to have the possibility to use global tests rather than local ones.

\subsection{The evolution of the abundance of clusters}

The abundance of clusters at high redshift has been used as a cosmological 
constraint more than ten years ago (Peebles et al., 1989; Evrard, 1989). 
Ten years ago, Oukbir 
and Blanchard (1992) emphasized that the 
evolution of the abundance of clusters with redshift was rather different 
in low and high density universe,
 offering a possible new cosmological test. The interest of this test is 
that it 
is 
global, not local, and therefore allows to  actually constraint directly 
$\Omega_m $.
It is relatively insensitive to the cosmological constant.
In principle, this test  is relatively easy to apply, because  the  abundance 
at redshift $\sim 1.$ is more than an order of magnitude less in a critical 
universe, while it is almost constant in a low density universe. Therefore the
 measurement of the temperature distribution function (TDF) even at 
$z \sim 0.5$ should
 provide a robust answer.
In recent years,  this test has received  considerable attention (Borgani et al, 1999; 
Eke et al., 1998; Henry, 1997; Henry, 2000; Viana and Liddle, 1999, among others). The first 
practical application  was by Donahue (1996) who emphasized that the 
properties of MS0451, with a temperature of around 10 keV at a redshift 
of 0.55, was already a serious 
piece of evidence in favor of a low density universe. This argument 
has been comforted
by the discovery of a high temperature cluster at 
redshift $z\sim 0.8$,
MS1054, which has a measured temperature of $\sim 12$ keV 
(Donahue et al, 2000). In the mean time, 
however, the redshift distribution of EMSS clusters was found to be well 
fitted by a high density universe under the assumption of a non evolving 
luminosity-temperature relation (Oukbir and Blanchard, 1997; Reichert et al., 
1999), as seems to follow from the properties of distant X-ray clusters
(Mushotsky, R.F. and Sharf, 1997; Sadat et al., 1998). 

 Application of this test is the purpose  of the XMM $\Omega$ program during the
 guaranty time phase (Bartlett et al., 2001).  In principle, this test can also be 
applied by using  other mass estimates, like velocity dispersion (Carlberg et al, 1997), 
Sunyaev-Zeldovich (Barbosa et al, 1996), or weak lensing. However, mass estimations based on X-ray 
temperatures is up to now the only method which can be applied at low and high 
redshift with relatively low systematic uncertainty. For instance, if velocity
 dispersions at high redshift ($\sim 0.5$) are overestimated by 30\%, the 
difference between low and high density universe is canceled. Weak lensing 
and SZ surveys of clusters to allow this test remain to be done.

\subsection{The local temperature distribution function}

In order to estimate the amount of evolution in the number of clusters, one 
obviously needs a reliable estimate of the number of clusters at $z \sim 0$. This already 
is 
not so easy and is a serious limitation. 
The estimation of the local temperature distribution function of X-ray clusters
 can be achieved from a sample of X-ray selected clusters for which the 
selection function is known,  and for which temperatures are available. Until 
recently, the standard reference sample was the Henry and Arnaud sample 
(1991), based on 25 clusters selected in the $2.-10.$ keV band. The ROSAT 
satellite has since provided  better quality samples of X-ray clusters, like 
the RASS and the BCS sample, containing several hundred of clusters. 
Temperature information is still lacking for most of clusters in these samples
 and therefore such clusters samples do not  allow yet to improve 
estimations of the TDF in practice. We have 
therefore constructed a sample of X-ray clusters, by selecting all X-ray 
clusters with a flux above $2.2 10^{-11}$ erg/s/cm$^2$ with $|b| > 20$. Most 
of the clusters come from the Abell XBACS sample, to which some non-Abell 
clusters were added. The completeness was estimated by comparison with the 
RASS and the BCS and found to be of  the order of 85\%. This sample comprises 50 
clusters, which makes it the largest one available for measuring the TDF
at the time it was published.  The inferred TDF is in very good agreement with
 the TDF derived 
from the BCS luminosity function or from more recent comprehensive survey (Reiprich and B\"ohringer, 2002) (with 
$\sim 65$ clusters). The abundance of clusters is higher than 
derived from the Henry and Arnaud sample as given by Eke et al. (1998) for 
instance. It is in good agreement with Markevitch (1998) for clusters with 
$T> 4$ keV, but is slightly higher for clusters with $T \sim 3$ keV. The power
 spectrum of fluctuations can be normalized from the abundance of clusters, 
leading to $\sigma_8 = \sigma_c = 0.6$ (using PS formula) for $\Omega_m = 1$ and
 to $\sigma_c = 0.7$
for $\Omega_m = 0.35$ corresponding to $\sigma_8 = 0.96$ for a $n = -1.5$ power 
spectrum index (contrary to a common mistake the cluster abundance does not 
provide an unique normalization for $\sigma_8 $ in low density models, but on a scale $\sim ^{-3}\sqrt{\Omega_m}8 h^{-1}$Mpc),  
consistent with recent estimates
 based on optical analysis of galaxy clusters (Girardi et al., 1998) and weak 
lensing measurements (Van Waerbeke et al., 2002).

\subsection{Application to the determination of $\Omega_m$}

The abundance of X-ray clusters at $z = 0.33$ can be determined from Henry 
sample (1997) containing 9 clusters. Despite the limited number of clusters 
and the limited range of redshift for which the above cosmological test can be
 applied, interesting answer can already be obtained, demonstrating the power 
of this 
test. Comparison of the local TDF and the high redshift TDF clearly show that 
there is a significant evolution in the abundance of X-ray clusters 
(see figure 1), such an evolution is unambiguously detected in our analysis. 
This evolution is consistent with the recent study of Donahue et al. (2000). 
We have 
performed a likelihood  analysis to estimate the mean density of the universe 
from the detected evolution between $z = 0.05$ and $z = 0.33$. The likelihood 
function is written in term of all the parameters entering in the problem: the
 power spectrum index and  the amplitude of the fluctuations. The best 
parameters
are estimated as those which maximize the likelihood function. The results 
show that for the open and flat cases,
one obtains  high values for the preferred $\Omega_m$ with a rather low error 
bar :
\begin{eqnarray}
\Omega_m = & 0.92^{+0.26}_{-0.22}   &  \mbox{  {\textrm (open case)}}\\
 \Omega_m = & 0.86^{+0.35}_{-0.25}     & \mbox{ {\textrm (flat case)}}
\end{eqnarray}
(Blanchard et al., 2000). Interestingly, the best fitting model also reproduces the abundance of 
clusters (with $T \sim 6$ keV) at  $z = 0.55$ as found by Donahue and Voit
 (2000). 
 
\begin{figure}[ht]
\centerline{\epsfxsize=4.5in\epsfbox{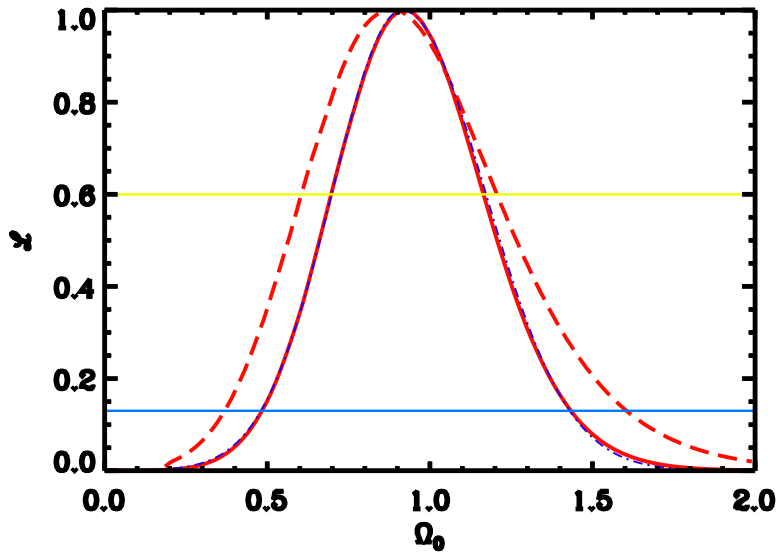}}   
\caption{ Likelihoods from the measured abundance of EMSS clusters in the redshift range (0.3,0.4) based on the Henry's sample (1997). The dashed line is for a flat universe while the continuous line is for an open cosmology.
\label{fig:likeli}}
\end{figure}

\subsection{ Systematic uncertainties in the determination of $\Omega_m$}

The above values differ sensitively from several recent analyzes on the same 
test and using the same high redshift sample. It is therefore important to 
identify the possible source of systematic uncertainty that may explain 
these differences. The test is based on the evolution of the mass function
(Blanchard and Bartlett, 1998). The mass function has to be related to the
primordial fluctuations. The Press and Schechter formalism is generally 
used for this, and this is what used in deriving the above numbers. 
However, this may be slightly 
uncertain. Using the more recent form proposed by Governato et al. (1999) 
we found a value for $\Omega_m$ slightly higher (a different mass function was used in 
Figure 1). A second problem lies in the mass temperature 
relation which is necessary to go from the mass function to the temperature 
distribution function. The mass can be estimated either from the hydrostatic 
equation or from numerical simulations. In general hydrostatic equation leads
to mass smaller than those found in numerical simulations (Roussel et al., 2000; 
Markevitch, 1998; Reiprich and B\"ohringer, 2002; Seljak, 2002).
 Using the 
two most extreme mass--temperature relations inferred from numerical 
simulations,
 we found a 10\% difference. We concluded that such uncertainties are not 
critical.  

An other serious issue is the local sample used: using HA sample 
 we found a value  smaller by $40\%$. Identically, if we postulated that the 
high redshift abundance has been underestimated by a factor of two, $\Omega_m$
is reduced by 40\%. The determination of the selection function of EMSS 
is therefore critical. An evolution in the morphology of clusters with
redshift  would result in a dramatic change in the inferred abundance (Adami et al., 
2001). This is the most serious possible uncertainty in this 
analysis. However, the growing evidences for the scaling of observed 
properties of distant clusters  (Neumann and Arnaud, 2001),  rather 
disfavor such possibility.

\subsection{An other global test : the baryon fraction in local clusters}

This is a very interesting test proposed by White et al. (1993) which in principle offer a rather direct way 
to measure $\Omega_m$. It relies on one side on the fact that one should be 
able to 
measure the total mass of clusters, as well as their baryon content and on the
other side that the primordial abundance of baryons can be well constrained 
from the predictions of primordial nucleosynthesis and the observed abundances 
of light elements. Furthermore, the CMB is providing interesting constraints 
on the baryon density of the universe, that are essentially consistent with values inferred 
from nucleosynthesis (Eq. \ref{eq:obbn}).
X-ray observations of clusters allow to measure their gas mass which 
represents the dominant component of their (visible) baryonic  content 
(the stellar component represents around 1\% of the total mass). In this way 
one can measure the baryon fraction $f_b$ and infer $\Omega_m$:
$$\Omega_m = \gamma^{-1}\frac{\Omega_{bbn}}{f_b}$$
where $\gamma $ represents a correction factor between the actual baryon
fraction and the naive value ${\Omega_{bbn}}/{\Omega_m}$; 
typically,  $\gamma \sim 0.9 $. This method has been 
used quite often (Evrard, 1997; Roussel et al., 2000). There are some differences between measurements, mainly due 
to the mass estimators used. One key point is that the baryon fraction has to be 
estimated in the outer part of  clusters as close as possible to the
virial radius. However,
the outer profile of the X-ray gas has been shown by Vikhlinin et al. 
(1999) not to follow the classical
 $\beta$ profile, usually assumed, but being actually steeper;
   consequently derived gas masses are somewhat lower than from usual 
analysis. Recently, several consequences of this work 
were derived on the baryon fraction (Sadat and Blanchard, 2000):
\begin{itemize}
\item the scaled baryon fraction flattens in the outer part of clusters.
\item the global shape of the baryon fraction from the inner part to the 
outer part follows rather closely the shape found in numerical simulations
from the Santa Barbara cluster project (Frenk et al, 1999). 
\item when mass estimates are taken  from numerical simulations the 
baryon fraction, corrected from the --rather uncertain-- clumping  factor 
(Mathiesen et al, 1999) 
could be as low as 10\% ($h = 0.5$).
\end{itemize}

 The consequence of this is that a value of $\Omega_m$ as high as $0.8$ can be acceptable.
Large systematic uncertainties are still possible, and value twice lower can 
certainly not be rejected on the basis of this argument, but similarly
a value $\Omega_m \sim 1$ can not be securely rejected.

\subsection{The  baryon fraction in high redshift clusters}

\begin{figure}
\centerline{\epsfxsize=4.in\epsfbox{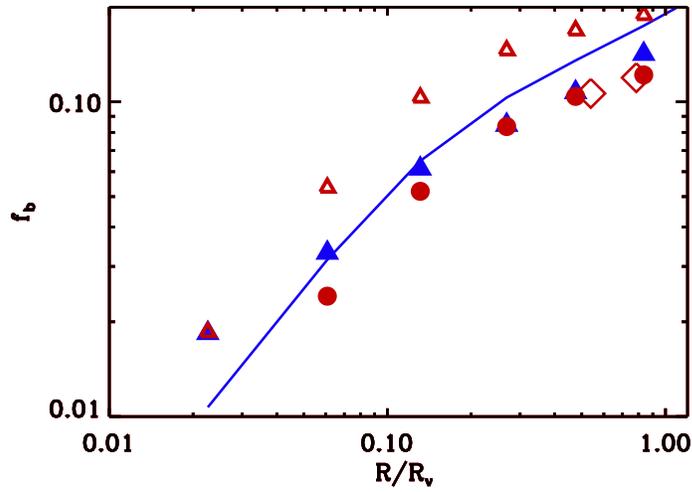}}   
\caption{ \label{fig:rxj} From the observed X-ray surface brightness of the distant 
cluster RXJ1120 (Arnaud et al., 2002) the gas fraction density profile 
(red filled circles) is compared to the results from the local clusters derived by 
Roussel et al. (2000) (blue open triangles) and those found in the outer regions
 by Sadat and Blanchard (2000) (red rhombuses). The profile 
shape is very close to those of local clusters. The amplitude is right for 
an $\Omega_m = 1.$ model, while a lambda model (open red triangles) is 
in strong disagreement with the data.
}
\end{figure}

A reasonable assumption is that the baryon fraction in clusters should remain
 more or less constant with redshift, as there is no motivation for 
introducing a variation with time of this quantity. When one infers the baryon 
fraction from X-ray observations of  clusters at cosmological distances,
the background cosmology is coming in the inferred value, through angular and
luminosity distances. Therefore for a given observed cluster, the inferred 
gas fraction would vary accordingly to the cosmology. This opens a way to 
constraint the cosmology, if one assumes that the apparent 
baryon fraction has to be constant (Sasaki, 1996; Pen, 1997), 
or equivalently that the emissivity 
profiles of clusters has to be identical when  scaling laws are taken into 
account (Neumann and Arnaud, 2001). Application of this test probably needs a 
large statistical  sample, but a preliminary application can be done on a
 distant cluster observed by XMM: RXJ1120.

This distant  cluster is a perfect 
candidate for the application of this test: 
the X-ray emission has been detected up to a distance close to the virial 
radius (Arnaud et al., 2001), the cluster is a $\sim 6$ keV cluster, with a relaxed configuration.
The gas profile can be derived up to a radius of the order of the virial 
radius without extrapolation. The inferred radial gas profile possesses two remarkable properties:
i) the shape of the gas profile in this distant cluster is in very good 
agreement with the shape of the profile inferred from local clusters by Sadat and Blanchard (2000), giving 
an interesting further piece of evidence in favor of this shape ii) the amplitude matches the amplitude of 
the local sample only for a high matter density universe, while an universe 
dominated by a  cosmological constant is strongly disfavored.

\section{Conclusion}

In this paper I have presented a personal point of view on the observational determination of cosmological 
parameters and especially on  question of the possible non-zero value of 
the cosmological constant. 
Although, the concordance model provides a nice agreement with    
several observational data sets, I have argued that i) the only direct case foran accelerating universe, implying the 
domination of the vacuum density over the other type of dark matter
already assumed to be  present in the Universe 
(baryonic dark matter, non-baryonic dark matter), is  coming from the distant 
SNIa and is not sufficient to be considered as robustly established. ii) some 
evidences against the concordance model are systematically rejected, because 
they are judged as insufficiently robust. The  global picture 
drawn by the concordance model might be right 
after all! But I still consider that  the case for a cosmological constant 
is oversold. It would be crucial in order to strength the case to have 
independent evidence either direct or indirect. A possible way for this would 
be to achieve a reliable measurement of the matter density of the Universe,
which in conjunction with the CMB evidence for flatness, would allow an 
estimate of the cosmological constant.
I have argued that clusters are in several ways the best tool 
to achieve such a measurement. Again contrary to a common prejudice I have 
illustrated that there are different values obtained by such methods, 
{\em some corresponding to high matter density} consistent with an Einstein-de Sitter model.\\

Summarizing results on clusters, I have shown an up-to-date local temperature 
distribution function obtained from a flux limited ROSAT sample comprising
 fifty clusters. When  compared to  Henry's sample at 
$z = 0.33$, obtained from the EMSS,   this sample clearly indicates 
that the TDF is 
evolving. This evolution is consistent with the evolution detected 
up to redshift $z = 0.55$ by Donahue et al. (1999).  This indicates converging
 evidences for a high density universe, with a value of $\Omega_m$ consistent 
with what Sadat et al. (1998) inferred previously from the full EMSS sample
taking into account the observed  evolution in the $L_x-T_x$ relation (which 
was found  moderately positive and consistent with no evolution).
From such analyzes, low density universes with $\Omega_m \leq 0.35$ 
are excluded at the two-sigma 
level. This conflicts with some of the previous analyzes on the same high 
redshift sample. Actually, lower values obtained from statistical analysis 
of X-ray samples were primarily
 affected by the biases introduced by the local reference sample,
which lead to a lower local abundance and a flatter spectrum for primordial 
fluctuations 
(Henry, 1997, 2000; Eke et al., 1998; Donahue and Voit, 1999). Our result is 
consistent with the conclusion of Viana and Liddle (1999), 
Reichert et al. (1999) and Sadat et al 
(1988). The possible 
existence of high temperature clusters at high redshift, MS0451 (10 keV) 
and MS1054 (12 keV), cannot however be made consistent with this picture 
of a high density universe, 
unless their temperatures are overestimated by a large factor  or the 
primordial fluctuations are not gaussian. The baryon fraction in clusters 
is an other global test of $\Omega_m$, provided that a reliable value for 
$\Omega_b$ is obtained. However, it seems that 
the mean baryon fraction could have been 
overestimated in previous analysis, possibly 
being closer to 10\% rather than to 
15\%-25\%. This is again consistent with a high density universe.
Finally, we have seen in one case that the apparent evolution of the baryon 
fraction in clusters could also be consistent with a high density universe. \\

In conclusion, I pretend that the determination of cosmological parameters
and especially the evidence for a non-zero cosmological constant is still an 
open question which needs to be comforted and that the exclusion of an 
Einstein de Sitter model is over-emphasized.

\end{document}